\documentclass[acmtog]{acmart}
\AtBeginDocument{%
  }

\copyrightyear{2025}
\acmYear{2025}
\acmConference[SA Conference Papers '25]{SIGGRAPH Asia 2025 Conference Papers}{December 15--18, 2025}{Hong Kong, Hong Kong}
\acmBooktitle{SIGGRAPH Asia 2025 Conference Papers (SA Conference Papers '25), December 15--18, 2025, Hong Kong, Hong Kong}\acmDOI{10.1145/3757377.3763950}
\acmISBN{979-8-4007-2137-3/2025/12}

\begin{document}

\title{Motion In-Betweening for Densely Interacting Characters}

\author{Xiaotang Zhang}
\email{xiaotang.zhang@durham.ac.uk}
\orcid{0000-0003-0822-9064}
\affiliation{%
  \institution{Durham University}
  \city{Durham}
  \country{United Kingdom}
}

\author{Ziyi Chang}
\email{ziyi.chang@durham.ac.uk}
\orcid{0000-0003-0746-6826}
\affiliation{%
  \institution{Durham University}
  \city{Durham}
  \country{United Kingdom}
}

\author{Qianhui Men}
\email{qianhui.men@bristol.ac.uk}
\orcid{0000-0002-0059-5484}
\affiliation{%
  \institution{University of Bristol}
  \city{Bristol}
  \country{United Kingdom}
}

\author{Hubert P. H. Shum}
\authornote{Corresponding Author.}
\email{hubert.shum@durham.ac.uk}
\orcid{0000-0001-5651-6039}
\affiliation{%
  \institution{Durham University}
  \city{Durham}
  \country{United Kingdom}
}

\renewcommand{\shortauthors}{Zhang et al.}

\begin{abstract}
Motion in-betweening is the problem to synthesize movement between keyposes. Traditional research focused primarily on single characters. Extending them to densely interacting characters is highly challenging, as it demands precise spatial-temporal correspondence between the characters to maintain the interaction, while creating natural transitions towards predefined keyposes. In this research, we present a method for long-horizon interaction in-betweening that enables two characters to engage and respond to one another naturally. 
To effectively represent and synthesize interactions, we propose a novel solution called Cross-Space In-Betweening, which models the interactions of each character across different conditioning representation spaces. 
We further observe that the significantly increased constraints in interacting characters heavily limit the solution space, leading to degraded motion quality and diminished interaction over time. 
To enable long-horizon synthesis, we present two solutions to maintain long-term interaction and motion quality, thereby keeping synthesis in the stable region of the solution space.
We first sustain interaction quality by identifying periodic interaction patterns through adversarial learning. 
We further maintain the motion quality by learning to refine the drifted latent space and prevent pose error accumulation. 
We demonstrate that our approach produces realistic, controllable, and long-horizon in-between motions of two characters with dynamic boxing and dancing actions across multiple keyposes, supported by extensive quantitative evaluations and user studies.
\end{abstract}

%
%



\begin{teaserfigure}
\includegraphics[width=\textwidth]{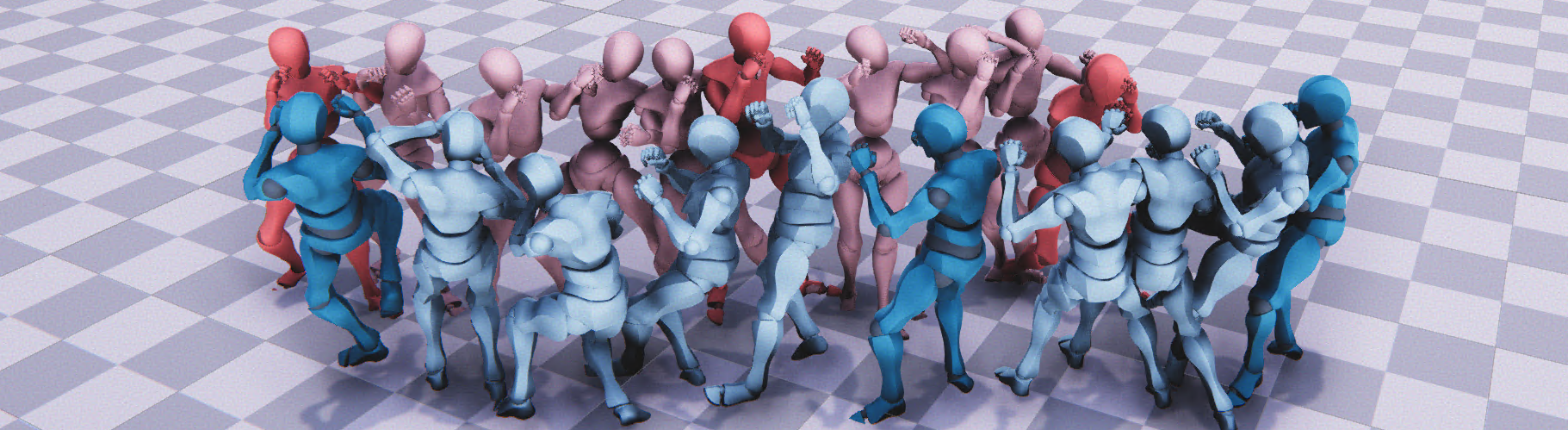}
\caption{Our approach is capable of producing interactive in-between motions (light blue and pink characters) for multiple keyposes (blue and red characters) while maintaining high quality over long transitions.} 
\label{fig:Teaser}
\end{teaserfigure}


\maketitle 

\section{Introduction}
Motion in-betweening involves synthesizing realistic character movement between predefined keyposes. 
It enables animators to efficiently create controllable motions by specifying only keyposes. 
Prior studies \cite{studer2024factorized,hong_long-term_2024,chu_real-time_2024,chai2024dynamic,oreshkin2023motion,tang2022real,qin2022motion,harvey2018recurrent,harvey2020robust} have explored various architectures for the motion in-betweening task, under diverse conditioning schemes such as text \cite{pinyoanuntapong_mmm_2024,cohan2024flexible}, style \cite{tang2023rsmt}, skeletal topology \cite{yun_anymole_2025,gat_anytop_2025} and keyframe timing \cite{,starke2023motion,goel_generative_2025}.
Despite extensive research, existing methods primarily focus on motion in-betweening for a single character, and it is non-trivial to extend these methods to multiple densely interacting characters. 

Dense interactions such as boxing or dancing are characterized by precise movements as well as precise timing. 
Extending interaction synthesis to in-betweening requires generation to fulfill three constraints: 
(1) The two-character motion should be spatio-temporally aligned such that it’s semantically interactive. 
(2) The two-character motion should end at the predefined keypose at the same time.
(3) In-betweening should generalize to user-customized keypose which typically lies outside the spatio-temporal distribution learned from dataset.
These precise requirements introduce significantly more constraints to the solution space than single-character in-betweening.

Fundamentally, the core challenge of interaction in-betweening is two-fold. 
First, it requires explicitly modeling interactions to capture precise movement and timing that define the interaction, while dynamically satisfying keyposes for each character. 
Second, enforcing dense interaction introduces a substantial number of constraints. To fulfill them would easily lead to unnatural motion and cause keyposes unreachable. This degradation compounds over time, ultimately making long-horizon interaction synthesis infeasible.

In this paper, we present a novel solution for long-horizon, densely interacting in-betweening that enables two characters to engage and respond to one another naturally. 
To represent interaction effectively, we introduce \textit{Cross-Space In-Betweening} to model and synthesize reactive in-between motions for two characters.
Our approach first represents motions as spatial offsets relative to keyposes and synthesizes transitions for each character individually.
The transitions are then transformed relative to the other character, with interaction conditions integrated via an affine transformation learned by Feature-wise Linear Modulation (FiLM) \cite{perez2018film}. 
This two-stage synthesis ensures stable, responsive motion transitions conditioned on the relative positions to both the keyposes and the counterpart character.

To address the challenge of overly restrictive constraints in this task, we present two solutions that help to preserve interaction consistency and motion quality over time, enabling long-horizon interaction in-betweening.
We first sustain interaction quality by identifying periodic interaction patterns through adversarial learning, which distinguishes real temporal structures from synthetic, inconsistent interactions.
This design is inspired by the observation that dense interactions like boxing and dancing often follow periodic and repetitive spatial distance.
%
Second, we maintain individual motion quality by sampling from a refined latent space during inference—a simple yet effective strategy to mitigate error accumulation and distribution drift common in auto-regressive methods.
Together, these two strategies foster a robust latent space informed by interaction periodicity and individual motion correction, supporting our goal of high-quality long-horizon synthesis.

We showcase long-horizon interaction in-betweening across the Boxing \cite{shum2010simulating}, ReMoCap \cite{ghosh2025remos} and InterHuman \cite{liang2024intergen} datasets.
Our system enables users to interactively select, translate, and rotate keyposes for two characters, with valid interactions automatically generated in response (see Fig.\ref{fig:ControlKeypose}).
Ablation studies, quantitative evaluations, and user studies demonstrate that our method outperforms prior work on interaction in-betweening.

The main contributions of this work can be summarized as:
\begin{itemize}
    \item We propose Cross-Space In-betweening that enables stable and responsive interaction modeling for two-character motion in-betweening.
    \item We maintain long-term interaction quality by identifying periodic interaction patterns through adversarial learning.
    \item We preserve long-term motion quality by learning to refine the drifted latent space and prevent pose error accumulation.
    \item We demonstrate that our system is robust to produce responsive in-between interactions for user-defined keyposes.
\end{itemize}

\begin{figure*}[h]
    \centering
    \includegraphics[width=\linewidth]{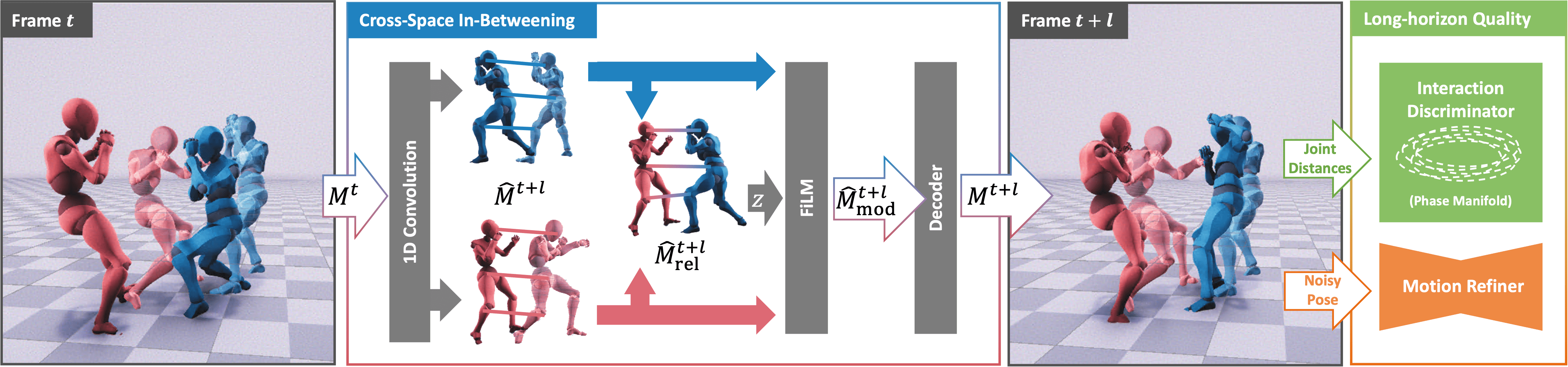}
    \caption{An overview of our framework. The system first generates an initial prediction for individual character which minimizes the distance to keypose. Then, it extracts relative pose representations as conditions to refine the initial prediction and generates interactive motions. Pairwise joint distances and the outcomes of main network are fed into an interaction discriminator and a motion refiner to model interaction periodicity and to reduce pose error, respectively.} 
    \label{fig:Architecture}
\end{figure*}

\section{Related Works}


\subsection{Multi-Character Interaction Synthesis}
Modeling interactions between virtual characters has been widely explored in computer graphics and vision. Early works relied on handcrafted patches or probabilistic models to capture interaction dynamics \cite{shum2008interaction,yun2012two,park2004analysis,kwon2008two,ho2009character,shen2019interaction}, but lacked flexibility and scalability. Deep neural networks have become the dominant paradigm in recent years. A number of works focus on predicting the short-term future of interacting characters \cite{katircioglu2021dyadic,guo2022multi,chopin2023interaction,tanke2023social}. These methods typically extrapolate trajectories from recent motion history. However, their scope is limited to short temporal horizons, and they generally lack mechanisms for conditional generation.
Another line of work generates the motion of one character conditioned on the observed trajectory of another. Recent approaches \cite{xu2024regennet,ghosh2025remos,cen2025ready_to_react} demonstrate reactive motion synthesis that align with the given input character’s ground-truth motion. However, this setting requires complete observation of motion as input, which makes it unsuitable for interaction in-betweening, where only sparse keyposes are available and both characters must be synthesized jointly.

Recently, diffusion-based models have shown promising results in multi-character generation \cite{tanaka2023role,xu2024regennet,shafir2023human,liang2024intergen,xu2023actformer}. These approaches typically employ text prompts or action labels as control signals, which provides flexibility for generating diverse interaction scenarios. However, such high-level conditions do not allow precise control over the spatial-temporal details of the desired interaction. \cite{zhang2023simulation} demonstrated interaction synthesis conditioned on character morphology (e.g., body height), and \cite{cen2025ready_to_react} generated reactions based on sparse joint positions, but neither method allows explicit control over the exact spatial-temporal interaction between characters. To bridge this gap, we propose a framework that allows two characters to perform natural interactions with guaranteed alignment to user-specified keyposes.

\subsection{Motion In-betweening}

Motion in-betweening has been extensively explored for single-character animation, ranging from early space-time optimization and probabilistic models \cite{ngo1993spacetime,rose1996efficient,witkin1988spacetime,wang2007gaussian,lehrmann2014efficient} to recent deep learning approaches using recurrent neural networks \cite{harvey2018recurrent,harvey2020robust}, Transformers \cite{chai2024dynamic,oreshkin2023motion,kim2022conditional}, mixture-of-experts \cite{tang2023rsmt,tang2022real,starke2023motion}, and diffusion models \cite{cohan2024flexible,studer2024factorized}. These methods achieve strong results for smooth, controllable single-character transitions under diverse conditioning schemes.

However, directly extending these approaches to two characters is non-trivial. Beyond matching individual keyposes, two-character in-betweening must explicitly model spatial relationships and timing alignment to maintain realistic interactions. Moreover, motion errors compound more rapidly in multi-character settings, where small deviations in one character can destabilize the interaction dynamics. Our method directly addresses these challenges by conditioning motion generation on cross-character spatial representations to preserve interaction fidelity and incorporating two solutions to maintain stable generation.

\section{Problem Definition}
\label{Problem Definition}
Our system is designed to auto-regressively predict future motion sequences for both characters based on an input observation sequence.
Specifically, the poses of both character $m_1, m_2 \in \mathbb{R}^{J \times C}$ are represented by $J$ joints and $C$ dimensions that capture joint positions and rotations. 
Given a motion dataset $\mathcal{M}$, we define the motion sequence of two characters starting at frame $t$ and spanning $T=20$ frames as 
$M^{t:t+T}=\{ (m^t_1,m^t_2), (m^{t+1}_1,m^{t+1}_2), \dots, (m^{t+T}_1,m^{t+T}_2) \}$
, which we denote as $M^{t}$ for simplicity. For each step of prediction, our framework predicts the short sequence of motion in the next $l=10$ frame based on past observation: 
${M}^{t+l} = f(M^{t})$. 
During inference, the in-between motion sequence is predicted in auto-regressive fashion until reaching the keypose.

\section{Methodology}
\subsection{Cross-Space In-betweening}
It is challenging to predict in-between motions in a two-character scenario, as the transitions pursue natural movements towards keyposes and simultaneously maintain high interaction quality. 
Thus, our approach strategically decomposes the problem into two key stages: \textit{individual in-betweening} and \textit{interaction modeling}.

In the first stage, individual motions are firstly represented in the coordinate space relative to the keypose ($M^{t}$) to obtain the spatial offset following the representation in \cite{starke2023motion}. An initial in-betweening prediction ($\hat{M}^{t+l}$) is then generated by minimizing the distance to the keypose.
In the second stage, the motion is then transformed into the coordinate space of its counterpart to obtain their relative spatial information ($\hat{M}^{t+l}_{\text{rel}}$) and to generate the final motion sequence ($M^{t+l}$). The architecture is shown in Fig.~\ref{fig:Architecture}. 

\subsubsection{Individual In-Betweening}
Specifically, we convert the input motion into frequencies using Discrete Cosine Transform (DCT) to effectively capture the temporal body dynamics and reduce modeling complexity.
An encoder (denoted as \textit{Enc}), composed of 1D convolutional layers followed by a Graph Convolutional Network (GCN) \cite{kipf2016semi} to capture the spatial dependencies among the joints, is then employed to predict the in-between motions for a single character.
This step is formulated as:
\begin{equation}
    \hat{M}^{t+l} = Enc(DCT(M^{t})).
\end{equation}

\subsubsection{Interaction Modeling}
In this stage, we convert the initial coarse prediction ($\hat{M}^{t+l}$) to the root space of the other character to obtain a new representation ($\hat{M}^{t+l}_{\text{rel}}$). 
These features indicate relative joints positions and rotations between the characters.
\begin{equation}
    \hat{M}^{t+l}_{\text{rel}} = \mathcal{T}(\hat{M}^{t+l}),
\end{equation}
where $\mathcal{T}$ denotes the spatial transformation applied to the motion representation to extract relative pose information.

Learning dependencies between motion features represented in different coordinate spaces is challenging due to they have significantly different distributions.
We thus incorporate Feature-wise Linear Modulation (FiLM) layers \cite{perez2018film} to adaptively bridge this feature gap. FiLM enables the network to modulate motion features through learnable affine transformations, conditioning each character's motion on spatial cues derived from its counterpart. This design is inspired by prior works in style conditioning \cite{mason_real-time_2022,tang2023rsmt}, where FiLM is used for feature modulation for different styles.

Specifically, we first project the relative-space motion features $\hat{M}^{t+l}_{\text{rel}}$ into a normally distributed latent space to regularize the interaction representation and stabilize FiLM conditioning \cite{kingma2013auto}.
Then, a FiLM layer is trained to condition the original keypose-space motions on the relative spatial information by modulating the motion features through learned affine parameters.
This process enables the model to align feature distributions across coordinate spaces while preserving interaction relevance.
The process can be formulated as:
\begin{equation}
    z = \mu(\hat{M}^{t+l}_{\text{rel}}) + \epsilon \cdot \sigma(\hat{M}^{t+l}_{\text{rel}}), \quad 
    \gamma, \beta = FiLM(z), 
\end{equation}
where $\mu$, $\sigma$ and $z$ are mean, log variance and re-parameterized latent variable following normal distribution, respectively. $\gamma$ and $\beta$ are learnable scale and shift parameters for motion feature modulation.

The FiLM layer integrates spatial conditions into the interaction modeling process by adaptively modulating the motion features. This modulation yields intermediate features, denoted as $\hat{M}^{t+l}_{\text{mod}}$, which is embedded with relative spatial information between characters. 
Subsequently, the decoder ($Dec$)—which shares the encoder’s architecture—reconstructs the complete in-between motion sequence based on $\hat{M}^{t+l}_{\text{mod}}$, followed by Inverse Discrete Cosine Transform (IDCT) to recover the final motion frames in the original pose space:
\begin{equation}
    \hat{M}^{t+l}_{\text{mod}} = \hat{M}^{t+l}\cdot \gamma + \beta, \quad
    M^{t+l} = IDCT(Dec(\hat{M}^{t+l}_{\text{mod}})).
\end{equation}


\subsubsection{Training}
During training, the outputs are fed back into the network as inputs in auto-regressive manner to enable sequential generation, with scheduled sampling \cite{bengio2015scheduled} adopted to further improve model robustness.
The network is optimized through minimizing mean squared error $\mathcal{L}_{\text{mse}}$ between prediction $M^{t+l}$ and ground truth $M^{t+l}_{\text{gt}}$, as well as a KL divergence loss $\mathcal{L}_{\text{kl}}$:
\begin{equation}
    \mathcal{L}_{\text{mse}} = \frac{1}{P} \sum^{P}
    (M^{t+l} - M^{t+l}_{\text{gt}})^2,
    \label{mse loss}
\end{equation}
\begin{equation}
    \mathcal{L}_{\text{kl}} = -0.5 \cdot \left(1 + \sigma - \mu^2 - e^{\sigma}\right),
\end{equation}
where $P = 3$ denotes the number of auto-regressive prediction steps performed during training.
The overall loss function for the Cross-Space In-betweening module is:
\begin{equation}
    \mathcal{L}_{\text{inbetween}} = 
    \lambda_{\text{mse}}\mathcal{L}_{\text{mse}} + 
    \lambda_{\text{kl}}\mathcal{L}_{\text{kl}}.
    \label{inbetween loss}
\end{equation}
Here, $\lambda_{\text{mse}}$, and $\lambda_{\text{kl}}$ are the corresponding weights that balance different losses.

\subsection{Long-horizon Quality}
We aim to generate interactive in-between motions that can extend over long horizon. 
However, the interaction in-betweening task introduces a substantial number of constraints.
These constraints can distort the structure of the learned latent space where valid trajectories become sparse and nonlinear. 
Inference-time distribution quickly drifts from the training one and pose error (e.g., deformed bones in Fig.~\ref{fig:PoseError}) accumulates after a few prediction steps.
Such disruptions in one character’s motion propagate abnormal features into the interaction modeling process, causing the network to generate diminished or overly smoothed interactions. (e.g., drifting to keypose without interactive motion, see Fig.~\ref{fig:WithoutInteractionDiscriminator}).
To address this challenge, we design two modules modeling interaction periodicity and enhancing motion quality, respectively, which helps to foster a robust latent space at both the interaction and single-character levels to enhance robustness of long-horizon synthesis.

\begin{figure}[h]
  \centering
  \includegraphics[width=\linewidth]{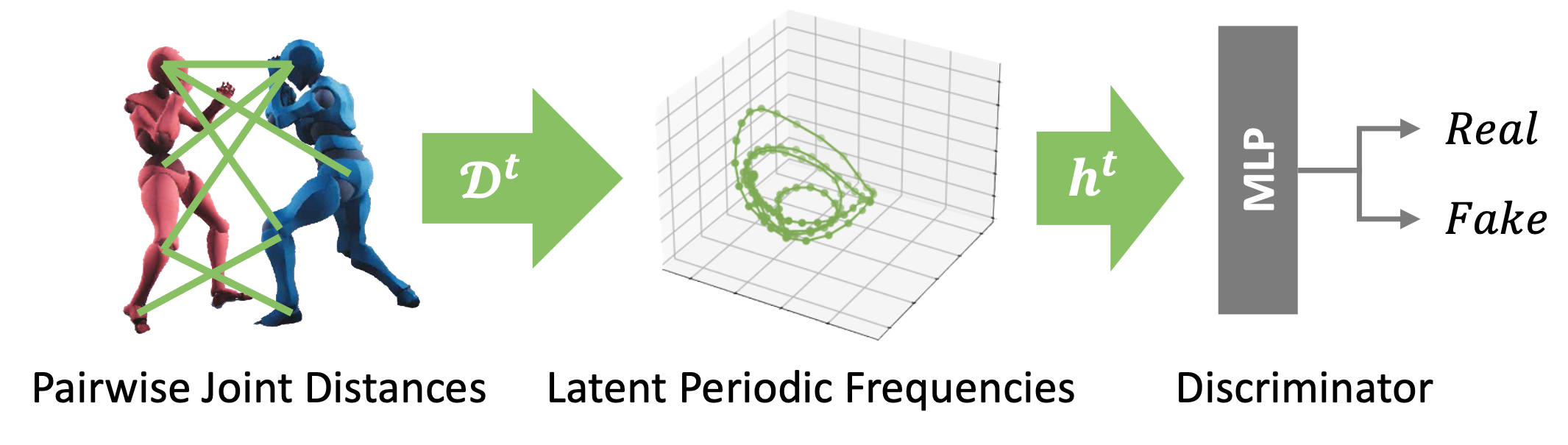}
  \caption{Details of the interaction periodicity modeling. We first extract the pairwise joint distances $\mathcal{D}^{t}$ from generated motions (green lines between pairs of joints). We then use Periodic Autoencoder to encode the dynamics as periodic latent frequencies $h^t$, illustrated as three principal components in the middle via principal component analysis. Our discriminator then learns to identify the interactions from periodic patterns between characters.}
  \label{fig:Interaction GAN}
\end{figure}

\begin{figure*}
  \centering
  \includegraphics[width=\linewidth]{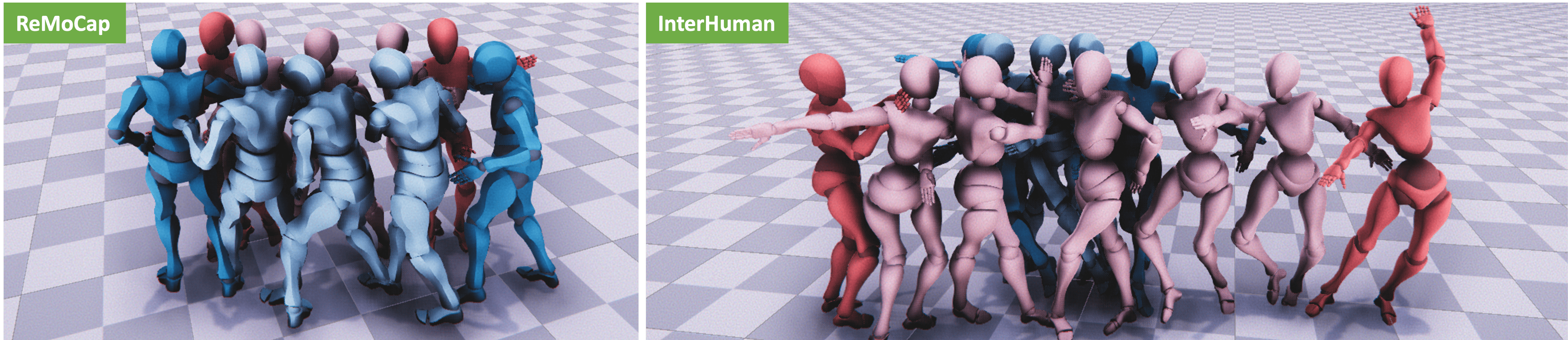}
  \caption{Qualitative results on ReMoCap and InterHuman dataset. Our method produces smooth and seamless turning motions (light blue and pink) in between keyposes (blue and red).}
  \label{fig:ReMoCapAndInterHuman}
\end{figure*}

\subsubsection{Interaction Periodicity}
Motivated by the observation that interactions like boxing and dancing exhibit inherent periodic patterns, we leveraged this property to enhance interaction quality and synchronization when generating in-betweening movements. 

Previous methods have modeled the periodic patterns of single character with the phase feature \cite{starke2022deepphase,starke2023motion,tang2023rsmt}, which enhance spatial-temporal alignment in the latent space in predicting subsequent poses. 
Similarly, we adopt Periodic Autoencoder (PAE)~\cite{starke2022deepphase} to learn the phase feature and generalize it to two characters to capture recurrent patterns in interactive actions by encoding their relative dynamics as periodic frequencies.
A discriminator is then deployed to evaluate the interaction quality of predicted motion sequences based on the learned periodicity. The procedure is shown in Fig.~\ref{fig:Interaction GAN}.

In two-character interactions, periodicity often emerges in the relational patterns between characters, rather than in the motions of each character independently. For example, repeated patterns occur as the punch approaches with decreasing distance between the characters, and then retracts with increasing distance. 
We thus model the interaction periodicity using pairwise joint distance (PJD) \cite{tang2008emulating}, which represents the geometry relationship between two characters. Here, we encode PJD dynamics as latent frequencies to represent the periodic patterns.
In particular, PJD is formulated as the per-frame offset of Euclidean distance of pairwise joints:
\begin{equation}
    d^{t} = \{ \| x^{t}_{i}-y^{t}_{j} \|^2_2 - \| x^{t-1}_{i}-y^{t-1}_{j} \|^2_2 \mid i,j \in J \},
\end{equation}
\begin{equation}
    \mathcal{D}^t = (d^{t}, d^{t+1}, \dots , d^{t+N}) \in \mathbb{R}^{J \times N},
    \label{pjd trajectories}
\end{equation}
where $x^{t}_{i}$,$y^{t}_{i}$$\in \mathbb{R}^{3}$ denote the joint positions of two characters in the world space at time step $t$, $d^{t}$ is the PJD at time $t$ for all joints, and $\mathcal{D}^{t}$ is PJD dynamics of length $N$ starting at time $t$.

The latent frequencies representation $h^{t} \in  \mathbb{R}^{N \times C_{\phi}}$ is parameterized by sinusoidal functions:
\begin{equation}
    h^{t} = PAE(\mathcal{D}^{t}) = a^{t}sin(2\pi (f^{t}+\phi^{t}))+b^{t},
\end{equation}
where $\phi^{t}$ is the phase vector predicted by a fully connected layer, and $f^{t}, a^{t}, b^{t}$ are the frequency, amplitude, and bias vector through Fast Fourier Transform (FFT) that models the cyclical dynamics of PJD between two characters. Phase channel $C_{\phi}$ is set to be 15.

To further increase the interaction realism, we train the main network (i.e., Cross-Space In-betweening) adversarially that learns to differentiate between realistic and unrealistic interactive patterns through the latent frequencies $h^{t}$. 
While the main network auto-regressively generates short-term motions, the discriminator offers global supervision by evaluating the periodic quality of the longer predictions (i.e., $N=30$ in Equation~\ref{pjd trajectories}). By identifying sampled sequences that could lead to erroneous interactions, it facilitates the generation of more realistic interactive motion patterns.

The adversarial loss $\mathcal{L}_{\text{adv}}$ is defined as:
\begin{equation}
\begin{split}
\mathcal{L}_{\text{adv}} = 
\mathbb{E}_{M^{t+l:t+l+N}_{\text{gt}} \sim \mathcal{M}}[\log D(M^{t+l:t+l+N}_{\text{gt}})] + \\
\mathbb{E}_{M^{t:t+N} \sim p_{G}}[\log(1 - D(G(M^{t:t+N})))]
\end{split}
\end{equation}
where $D$, $G$ are the discriminator and generator (i.e., Cross-Space In-betweening), $\mathcal{M}$ is the ground truth dataset and $p_{G}$ denotes the distribution of motion sequences generated by the generator $G$.

\subsubsection{Motion Quality}
Given that individual pose error will disrupt interaction consistency, we introduce a Motion Refiner to mitigate error accumulation and address motion degradation at the single-character level.
It learns to adjust the latent space derived from the drifted distribution generated by Cross-Space In-betweening and correct the deviations during inference so as to avoid sampling erroneous motions.

\begin{figure*}
  \centering
  \includegraphics[width=\linewidth]{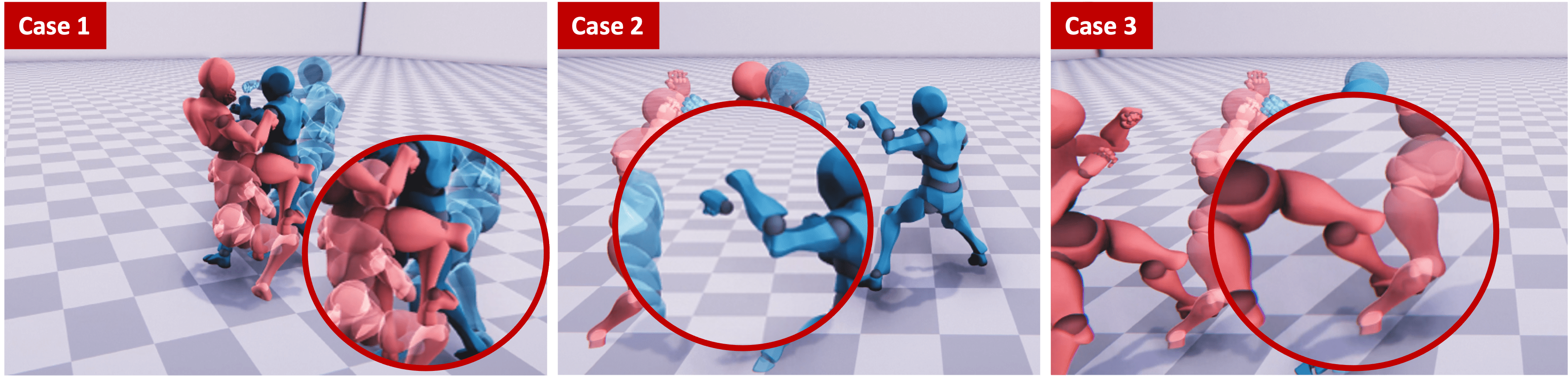}
  \caption{Examples of deformed bones caused by long-term error accumulation.}
  \label{fig:PoseError}
\end{figure*}

Specifically, the input to this module is the motion clip $M^{t+l}$ predicted by the main network, which may contain minor pose error (e.g. invalid joint rotations in Fig.~\ref{fig:PoseError}). 
Similar to Cross-Space In-betweening, the module consists of an auto-encoder and a GCN for spatial-temporal feature extraction and reconstruction. 
It preserves the high-level motion semantics of the input while refining joint-level features. The refined output is denoted as $M^{t+l}_{\text{refine}}$.
Empirically, we refine motion in non-overlapping segments of 10 frames that is sufficient for real-time inference while preserving motion variability. 


During inference, the Motion Refiner is integrated into the pipeline to correct motion predictions step-by-step.
This reduces the risk of accumulating pose errors or drifting into unrealistic latent spaces, and helps reshape the latent space to align with valid motion features.
In doing so, the refiner preserves long-term motion quality by avoiding dependence on the degraded distribution produced by the main network.
Note that the whole generation system is trained end-to-end, with the Motion Refiner updated independently:
\begin{equation}
    M^{t+l}_{\text{refine}} = Refiner(M^{t+l})
\end{equation}
\begin{equation}
    \mathcal{L}_{\text{refine}} = \frac{1}{P} \sum^{P}
    (M^{t+l}_{\text{refine}} - M^{t+l}_{\text{gt}})^2.
    \label{refinement loss}
\end{equation}

\section{Experiment}
\label{sec:Experiment}

\subsection{Implementation Details}
\paragraph{Simulation}
Our rendering system is implemented on top of an open-source motion animation framework \cite{starke2020local} developed in Unity3D. We do not adopt physics-based simulation as it introduces significant training overhead and requires extensive fine-tuning of torque and action smoothness \cite{liao2025beyondmimic}, which are not essential for demonstrating interactive motions in our setting.

\paragraph{Pose Representation}
To achieve efficient rendering, our system avoids using forward kinematics for joint position computation. Instead, each joint is represented independently in Cartesian 3D space, following \cite{starke2020local}.
Each character consists of $J=52$ joints and $C=9$ dimensions (3 for position and 6 for rotation) in total. 
Joint rotation is defined as a pair of forward and upward Cartesian vectors to avoid ambiguous rotations \cite{zhang2018mode}. 

\paragraph{Datasets}
We trained our model using a Boxing dataset as well as two public dancing motion datasets: ReMoCap \cite{ghosh2025remos} and InterHuman \cite{liang2024intergen}. Models are trained separately for each dataset.
The Boxing dataset is simulated and collected from a previous motion animation system \cite{shum2010simulating}, resulting in a total of 23,889 frames of intense boxing actions (e.g., punching, kicking and dodging).
For the ReMoCap and InterHuman datasets, we specifically select Lindy Hop and Latin dance sequences containing at least 500 frames and discard segments that lack clear interactions. This result in 39,557 and 18,224 total frames, respectively.
All datasets are re-targeted to Mixamo character \cite{Mixamo} using Autodesk Motion Builder \cite{autodesk_motionbuilder} and augmented by mirroring along the X-axis.
The dataset is divided into 90\% for training, 5\% for validation, and 5\% for testing. 

\paragraph{Keyframe Sampling}
Each training sample consists of 50 frames: the first 20 are used as reference input, and the remaining 30 for auto-regressive prediction.
During preprocessing, a keyframe is randomly selected between frames 50 and 70 (relative to the first frame). All 50 frames in the sample are then transformed into the coordinate space of this keyframe.
This approach ensures that the sampled keyframes cover both nearby and distant positions.

\paragraph{Training}
We conduct our experiments with an NVIDIA RTX 3080 graphics card, an AMD Ryzen 9 5900X CPU and 32G memory.
We train all modules using Adam Optimiser \cite{kingma2014adam} with a learning rate $\alpha=0.0001$, $\beta_{1}=0.9$, and $\beta_{2}=0.999$. 
Network is trained for 100 epochs and cost 40 to 50 hours for each dataset. 
The motion in-betweening runtime system is adapted from the open-source Unity3D motion animation system \cite{starke2020local}.

\subsection{Comparison Methods}
We aim to evaluate performance of the two objectives in our task, i.e., interaction synthesis and motion in-betweening, which were addressed separately in previous works. As there is no prior open-source work specifically targeting interaction in-betweening for direct comparison, we evaluate our system against the following representative baselines:
\begin{itemize}
    \item Phase Betweener \cite{starke2023motion}, a single-character in-betweening method with periodicity modeling. To evaluate its generalizability in a two-character setting, we input each character’s motion data either separately or concatenated, referred to as `separate' and `combined' in the following section.
    \item Cross-Interaction Attention \cite{guo2022multi}, a cross attention-based interaction synthesis method with explicit interaction modeling.
    \item CondMDI \cite{cohan2024flexible}, a diffusion-based single-character in-betweening framework.
\end{itemize}
Modification details for these baselines can be found in supplementary materials.

\begin{figure}
  \centering
  \includegraphics[width=\linewidth]{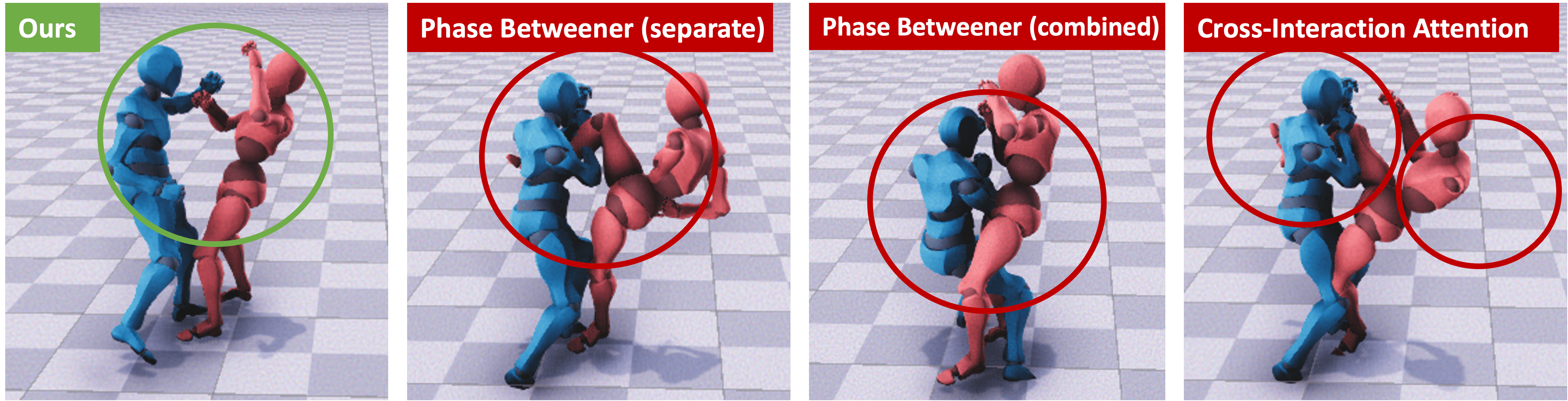}
  \caption{Qualitative results compared with baseline methods. Cross-Interaction Attention exhibits severe pose error accumulation issue.}
  \label{fig:Comparison}
\end{figure}

\begin{figure}
  \centering
  \includegraphics[width=\linewidth]{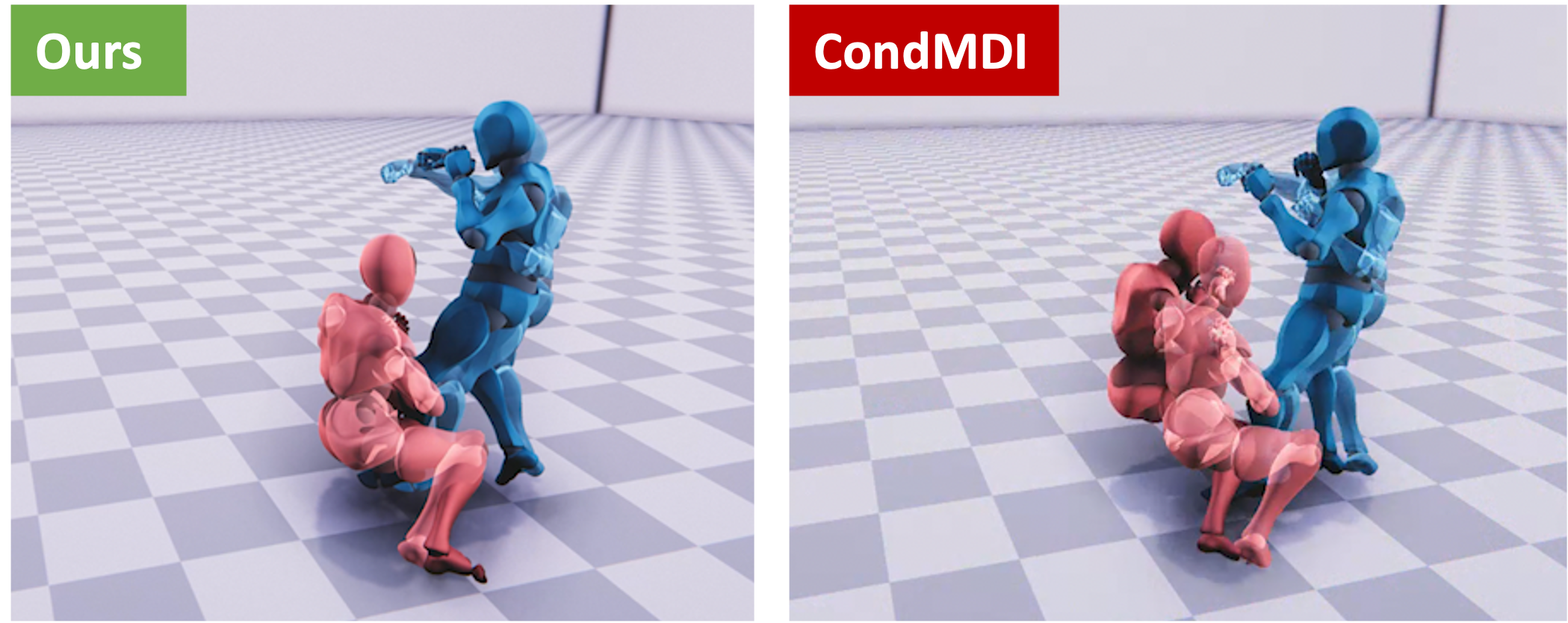}
  \caption{Keypose alignment performance compared with CondMDI.}
  \label{fig:CompareCondMDI}
\end{figure}

\subsection{Qualitative Results}
\begin{figure*}
  \centering
  \includegraphics[width=\linewidth]{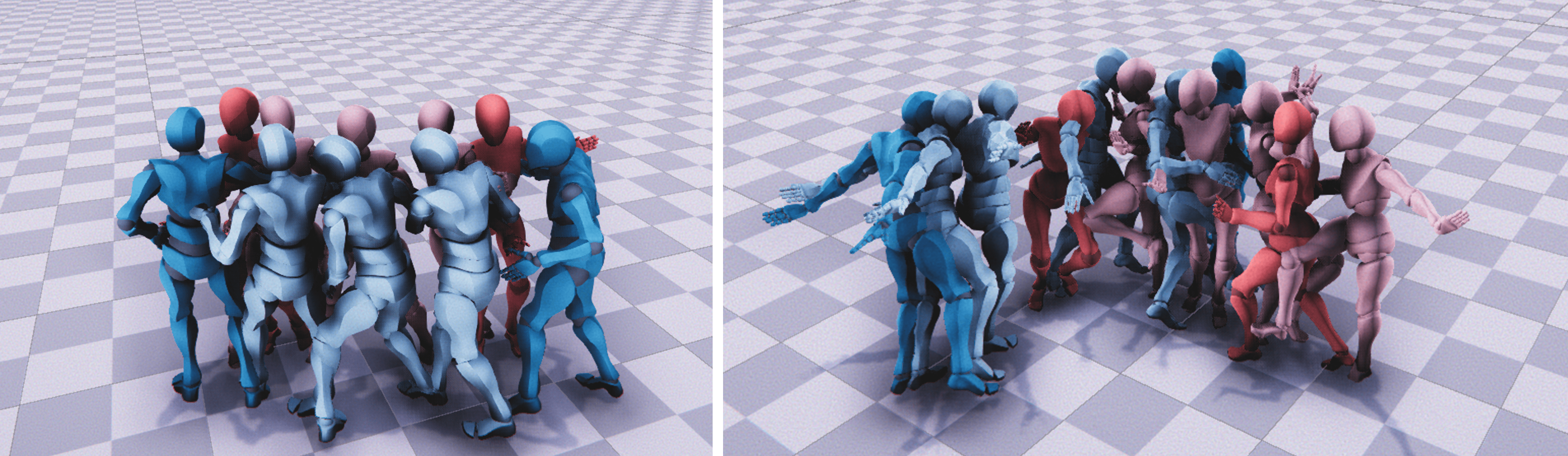}
  \caption{Qualitative results on ReMoCap dataset. Light blue and pink characters are in-between motion. Blue and red characters are keyposes.}
  \label{fig:ReMoCap}
\end{figure*}

\begin{figure*}
  \centering
  \includegraphics[width=\linewidth]{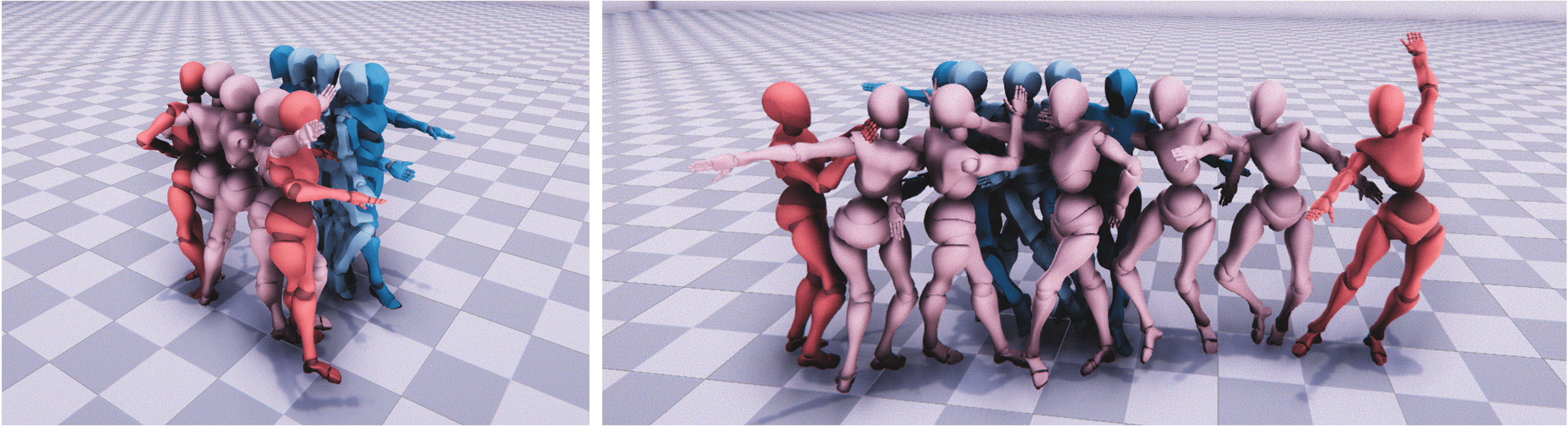}
  \caption{Qualitative results on InterHuman dataset. Light blue and pink characters are in-between motion. Blue and red characters are keyposes.}
  \label{fig:InterHuman}
\end{figure*}

\paragraph{Interaction In-betweening}
Our network is able to generate in-between motions in real time until the averaged distance between each character and its corresponding keypose falls below a predefined threshold. The inference time is 125 ms per 10 frames.
Qualitative results for Boxing, ReMoCap and InterHuman dataset can be found in Fig.~\ref{fig:Teaser},~\ref{fig:ReMoCap},~\ref{fig:InterHuman} and the supplementary video.
Comparisons with baselines are shown in Fig.~\ref{fig:Comparison}. 
For Phase Betweener, neither of training approaches yields realistic interactions, as the network fails to implicitly capture the complex relationships between characters.
Cross-Interaction Attention also fails to generate interactive in-between motions and suffers from severe error accumulation, as it relies heavily on historical motions for synthesis and struggles with long-term prediction.
Moreover, CondMDI struggles with keypose alignment (see Fig.~\ref{fig:CompareCondMDI}). Since it generates entire motion sequences in an offline manner, we loosen the keypose activation threshold in our real-time system to maintain continuity. This adjustment ensures smooth rendering but reduces alignment accuracy.

\paragraph{Controllability}
The keyposes are controllable by users.
We demonstrate its generalizability on different keyposes when provided with the same initial motion (See Fig.~\ref{fig:ControlKeypose} and 'Controllability' in supplementary video). 
Through customizing the root positions and rotations of the keyposes or sampling poses from database, our system is robust to produce responsive in-between interactive motion sequences in real time.

\begin{figure}
  \centering
  \includegraphics[width=\columnwidth]{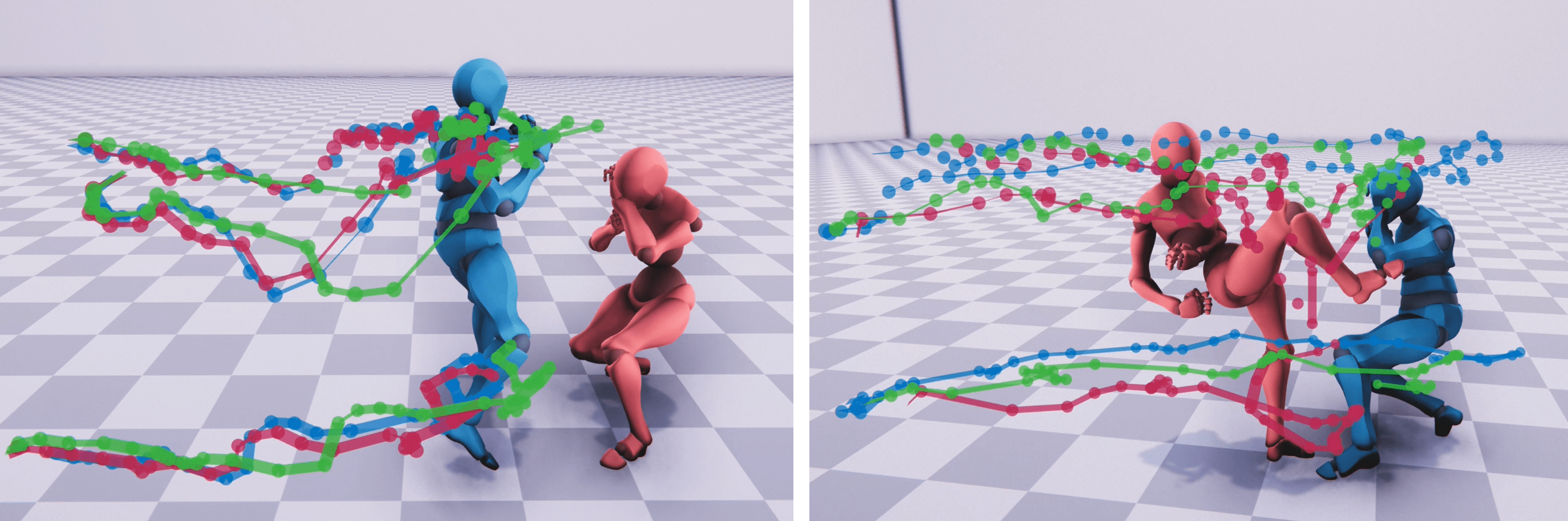}
  \caption{Qualitative results on predictions diversity. We illustrate historical trajectories in distinct colors for each of the three predictions.}
  \label{fig:Diversity}
\end{figure}

\paragraph{Diversity}
There are multiple possible transitions that can occur between keyframes. 
To demonstrate the prediction diversity, we generate three in-between motions for the same keypose condition, as shown in Fig.~\ref{fig:Diversity}.
Our network is capable of generating varied in-between motions while maintaining realistic interactions. 

\begin{table*}[!ht]
\caption{Quantitative results compared with previous methods and ablated versions. All comparison methods and ablated networks are trained on Boxing dataset only.}
\label{Quantitative Results}
    \centering
    \footnotesize
    \begin{tabular}{lcc|cc|c|ccc|cc|ccc|ccc}
    \toprule
        ~ & \multicolumn{2}{c|}{L2P$\downarrow$} & \multicolumn{2}{c|}{L2Q$\downarrow$} & \multicolumn{1}{c|}{Foot$\downarrow$} & \multicolumn{3}{c|}{Interaction$\uparrow$} & \multicolumn{2}{c|}{Diversity$\uparrow$} & \multicolumn{3}{c|}{FID$\downarrow$} & \multicolumn{3}{c}{100$\times$NPSS$\downarrow$} \\ 
        Frames & 30 & 50 & 30 & 50 & 50 & 40 & 60 & 80 & 30 & 50 & 100 & 150 & 200 & 100 & 150 & 200 \\
    \midrule
    \vspace{-0.7em} \\[-0.7em]
    \midrule
        \multicolumn{17}{c}{\textit{Boxing Dataset}} \\
        Ours &0.192&0.208&0.254&0.279&0.342&\textbf{0.914}&\textbf{0.911}&0.905&1.104&1.802&\textbf{0.282}&\textbf{0.287}&\textbf{0.289}&\textbf{1.398}&\textbf{1.430}&\textbf{1.531}\\ 
        w/o interaction modeling    &0.243&0.253&0.291&0.307&0.336&0.864&0.862&0.862&\textbf{1.112}&\textbf{1.831}&0.428&0.436&0.439&1.837&2.020&2.348\\ 
        w/o Inter. GAN              &\textbf{0.191}&\textbf{0.202}&\textbf{0.248}&0.266&0.340&0.904&0.902&0.899&1.090&1.787&0.504&0.529&0.556&2.016&2.306&2.551\\ 
        w/o Motion Refiner          &0.235&0.251&0.288&0.301&0.345&0.908&0.901&0.894&1.102&1.788&0.696&0.722&0.730&2.197&2.528&2.814\\ 
    \midrule
        Phase Betweener (separate)&0.211&0.214&0.257&0.263&0.345&0.866&0.863&0.859&1.089&1.670&0.519&0.534&0.565&2.119& 2.304& 2.663\\
        Phase Betweener (combined)&0.202&0.210&0.250&0.259&0.350&0.880&0.877&0.872&0.278&0.281&0.547&0.559&0.570&2.849&3.105&3.397\\
        Cross-Interaction   &0.217&0.218&0.267&0.272&0.474&0.882&0.875&0.871&0.314&0.317&0.482&0.493&0.502&2.147&2.335&2.590\\
        CondMDI & 0.212 & 0.217 & 0.263 & 0.266 & 0.452 & 0.905 & 0.902 & 0.900 & 0.293 & 0.306 & 0.478 & 0.484 & 0.496 & 1.833 & 1.940 & 2.056 \\
    \midrule
    \vspace{-0.7em} \\[-0.7em] 
    \midrule
    \multicolumn{17}{c}{\textit{Dancing Datasets}} \\
        Ours (ReMoCap Dataset)      &0.205&0.220&0.264&0.278&0.363&0.908&0.907&\textbf{0.907}&0.974&1.077&0.306&0.310&0.312&1.448&1.562& 1.793\\ 
        Ours (InterHuman Dataset)   &0.201&0.212&0.249&\textbf{0.259}&\textbf{0.335}&0.901&0.899&0.898&0.965&1.012&0.289&0.295&0.307&1.407& 1.494& 1.586\\ 
    \bottomrule
    \end{tabular}
\end{table*}

\subsection{Quantitative Results}
\paragraph{Reconstruction Quality}
To evaluate the reconstruction precision, we follow \cite{harvey2020robust} to report the average L2 norm of positions (L2P) and the average L2 norm of quaternions (L2Q) between the ground truth and the generated in-between motions in world coordinate, covering different lengths of in-betweening in short term (refer to Table~\ref{Quantitative Results}).
Our method demonstrates comparable performance across all three datasets and achieves superior reconstruction quality compared to baselines. 
Further results of long-term reconstruction error are provided in supplementary document.

\paragraph{Interaction Quality}
\label{InteractionMetric}
Inspired by \cite{wang2022discriminator}, we evaluate the quality of interactions based on the classification results of the discriminator. 
Note that the discriminator used for evaluation is trained independently on each of the three datasets using fake samples generated by a different network than the one being tested, so that to ensure an unbiased and effective assessment.
Our method outperforms the comparison baselines in terms of interaction quality. It also proves the effectiveness of using the PJD dynamics to represent and assess the interaction quality for periodic motions.

\paragraph{Long-horizon Quality}
To assess prediction quality in long horizon, in Table~\ref{Quantitative Results}, we also report the latent distribution differences with ground truth using Frechet Inception Distance (FID) and Normalized Power Spectrum Similarity (NPSS) \cite{harvey2020robust}. 
Our method shows robust performance in long transitions, but without the Motion Refiner it can produce highly unrealistic motions (as reflected in the FID metric). 
For NPSS, Phase Betweener (combined) performs the worst in long-horizon synthesis, likely due to the substantially increased complexity of learning a two-character state space with limited capacity of fully-connected layers.

\paragraph{Diversity}
We also report the diversity of predictions by measuring the joint positional difference between different samples given the same input and keypose condition. 
Similar to \cite{tang2023rsmt}, we generate 10 samples for each keypose condition. 
Our system achieves comparable performance to Phase Betweener. In contrast, Cross-Interaction Attention tends to be deterministic, as its attention mechanism generates similar attention scores for the same input.

\paragraph{Foot Sliding}
Following \cite{zhang2018mode}, we measure the foot sliding artifact (refer to \textit{Foot} in Table~\ref{Quantitative Results}) by calculating the averaged foot joint velocity $v_f$ in the first 50 frames when foot height $h_f$ is within threshold $H=2.5cm$: $v_f \cdot clamp(2-2^{h_f/H}, 0, 1)$. 
It is worth noting that the training datasets inherently contain some foot sliding artifacts. Following \cite{starke2020local}, we apply inverse kinematics on the foot joints to mitigate this issue. Quantitative results compared with ground truth are provided in supplementary document.

\begin{figure}[h]
  \centering
  \includegraphics[width=\linewidth]{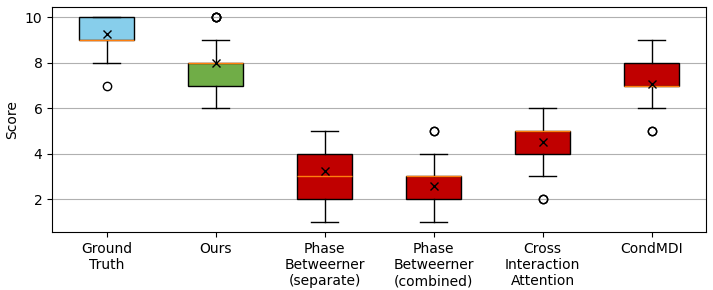}
  \caption{User study results presented as a box plot of ratings across different methods. Our method achieves scores comparable to the ground truth and surpasses all baseline approaches.}
  \label{fig:User Study}
\end{figure}

\paragraph{User Study}\
We conduct a user study with 50 participants who have no familiarity with motion in-betweening to assess the visual quality of the generated demos. 
Each participant is asked to rate the motion quality (on a scale of 1 to 10) for 4 demo cases (2 for boxing dataset and 2 for dancing datasets) generated by different methods, including ground truth. 
As shown in Fig.~\ref{fig:User Study}, our method consistently outperforms the baselines and achieves visual quality comparable to the ground-truth motions.
CondMDI achieved higher scores than the other baselines, which may be attributed to its ability to generate fewer unrealistic interactions and produce more continuous transitions across multiple keyposes.
Detailed statistics are included in the supplementary document.

\subsection{Ablation Studies}
\begin{figure}[t]
  \centering
    \centering
    \includegraphics[width=\linewidth]{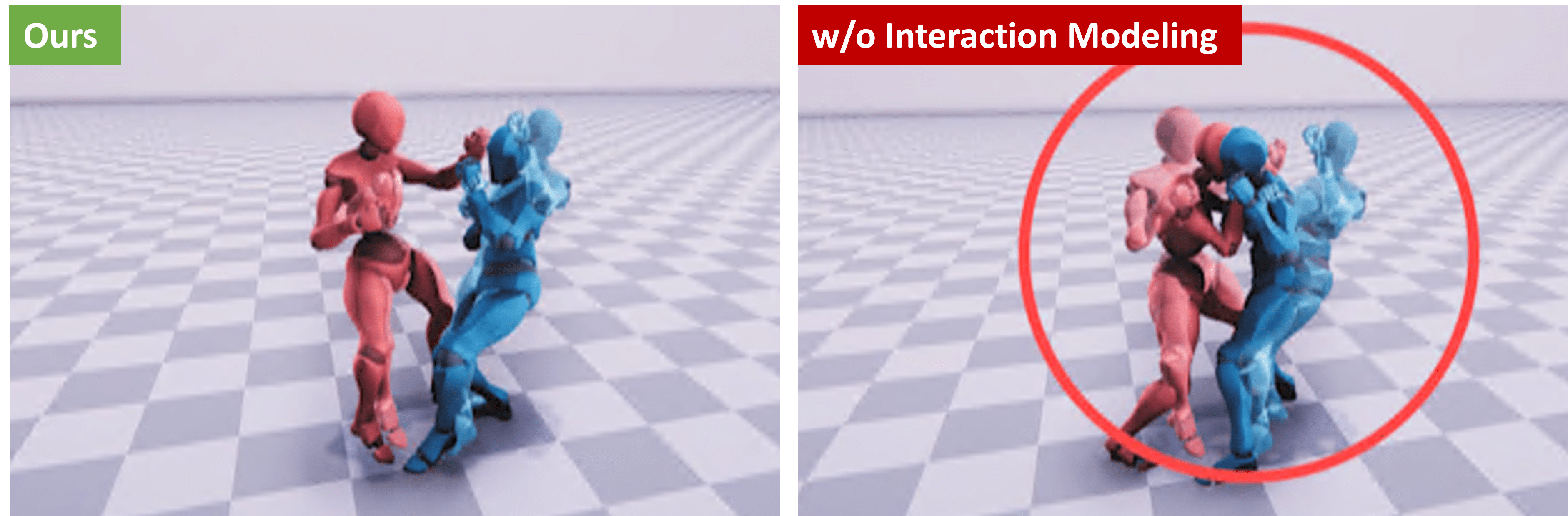}
    \caption{Qualitative comparison on interaction modeling. On the right, both characters exhibit significant penetration and unrealistic interactions.}
    \label{fig:WithoutInteractionModeling}
\end{figure}

\begin{figure*}[t]
    \centering
    \includegraphics[width=\linewidth]{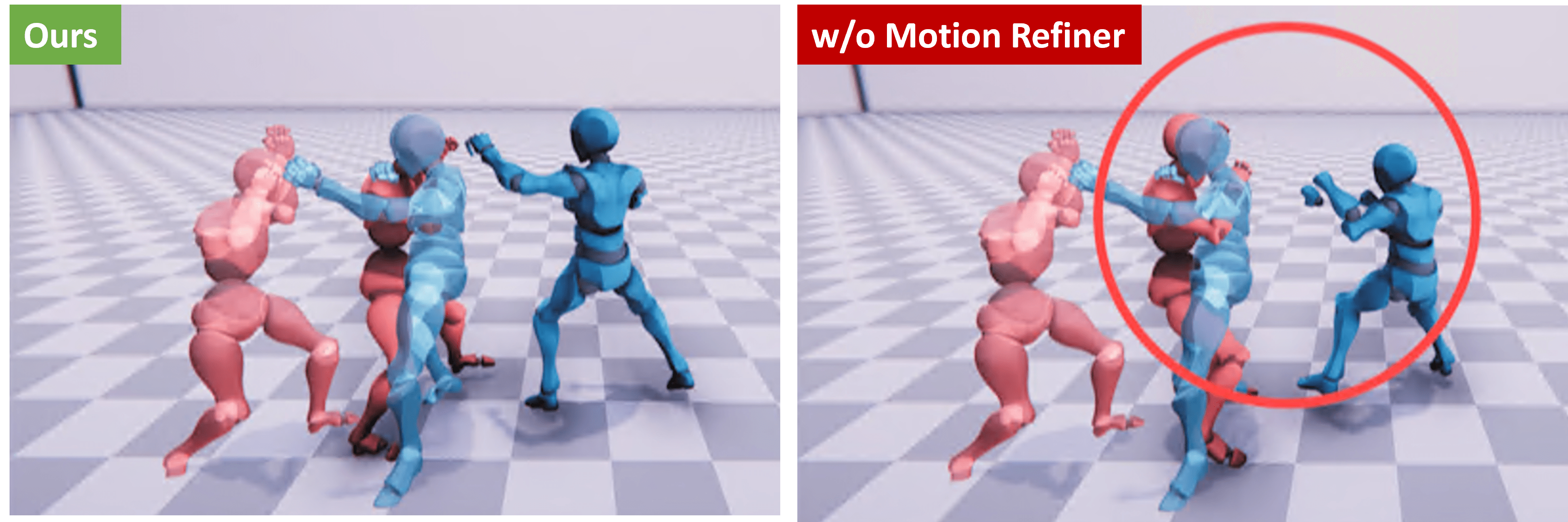}
    \caption{Qualitative results without the Motion Refiner. The blue character exhibits hand joint deformation after a few seconds of prediction.}
    \label{fig:WithoutMotionRefiner}
\end{figure*}

\begin{figure*}
  \centering
  \includegraphics[width=\linewidth]{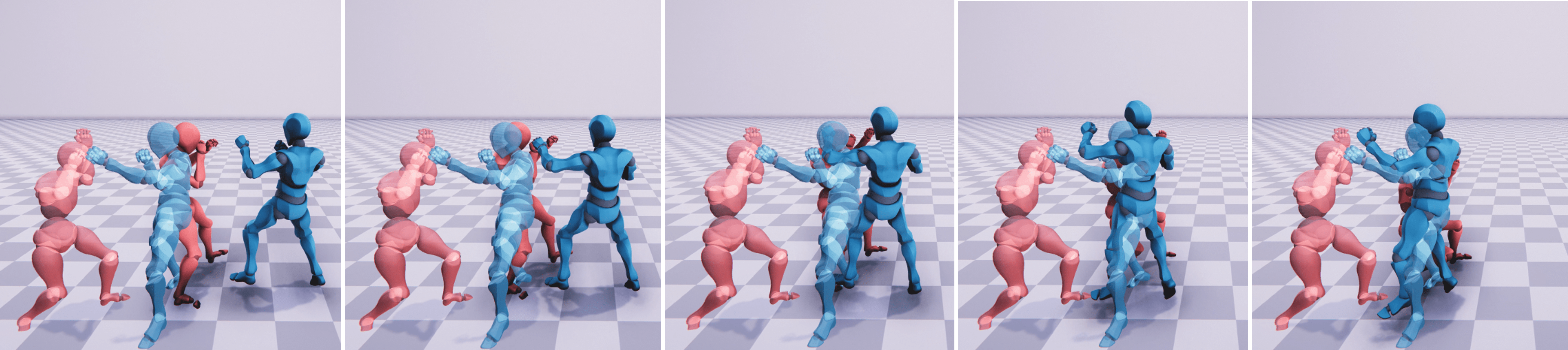}
  \caption{Qualitative results without modeling interaction periodicity. The blue character is sliding to its keypose without interactive movement.}
  \label{fig:WithoutInteractionDiscriminator}
\end{figure*}

We conduct ablation studies on Boxing dataset to evaluate the effectiveness of different components. 
To ensure fair comparisons and consistent model complexity, the ablated versions are implemented to have a similar number of parameters as the full model.
Qualitative results can be found in Fig.~\ref{fig:WithoutInteractionModeling},\ref{fig:WithoutInteractionDiscriminator},\ref{fig:WithoutMotionRefiner} and the supplementary video.

\paragraph{Cross-Space In-betweening}
To evaluate the necessity of FiLM-based interaction conditioning, we conduct an ablation (refer to \textit{w/o interaction modeling}) where each character's motion is predicted independently in its own keypose space without transforming to the other’s coordinate system or applying FiLM modulation. 
Quantitative results indicate a 20\% decline in reconstruction performance when dense spatial-temporal relationships between characters are not effectively captured. 

\paragraph{Interaction Periodicity}
We also test the interaction quality by ablating the discriminator (refer to \textit{w/o Inter. GAN} in the table).
We observe a minor improvement in short-term reconstruction precision, likely because removing the discriminator allows the network to focus directly on optimizing the reconstruction loss without the additional constraint of satisfying the adversarial loss.
However, we also observe a slight decrease in interaction quality and a significant decline in FID in long transitions. 
This reflects how modeling the interaction periodicity help learn a robust latent space and maintain long-horizon quality for in-betweening.

\paragraph{Motion Quality}
With the Motion Refiner, our network effectively maintain motion quality by reducing the pose errors in long-term synthesis, as shown in Fig.~\ref{fig:WithoutMotionRefiner} and supplementary video.
This is also evident from the lower discrepancy between latent distributions of ground truth and predictions in FID scores.

\section{Conclusion and Discussion}
In this work, we introduce Cross-Space In-betweening, a novel auto-regressive framework for synthesizing interactive in-between motions.
To address the challenges posed by strict keypose and interaction constraints, we maintain interaction and motion quality over long horizons, through modeling of interaction periodicity and refining individual pose errors.
Our system enables controllable, long-horizon interaction in-betweening with dense character interactions, and significantly outperforms existing methods across most quantitative metrics.

\paragraph{Limitation}
The first limitation of our work is that the motion may not match well with the user-customized keyposes because the imposed spatial constraints can be temporally misaligned (e.g., forcing two characters to punch simultaneously), and the model is unable to infer interactions it has not seen before.
Due to the nature of online synthesis, without a global motion planning mechanism, our system does not allow further offline refinement or adjustment of the entire generated in-between sequence for better keypose alignment.

Second, despite showcasing the potential of encoding the periodicity of two-character motions for improving the interaction quality, this strategy is primarily suited for interactions with clear repetitive patterns and does not readily extend to aperiodic interactions (see supplementary document).

Third, as the first work targeting interaction in-betweening, our system currently does not support in-between timing condition as it will further increase modeling complexity.


\begin{acks}
This research is supported in part by the EPSRC NortHFutures project (ref: EP/X031012/1).
\end{acks}

\bibliographystyle{ACM-Reference-Format}
\bibliography{references}

\begin{figure*}
  \centering
  \includegraphics[width=\linewidth]{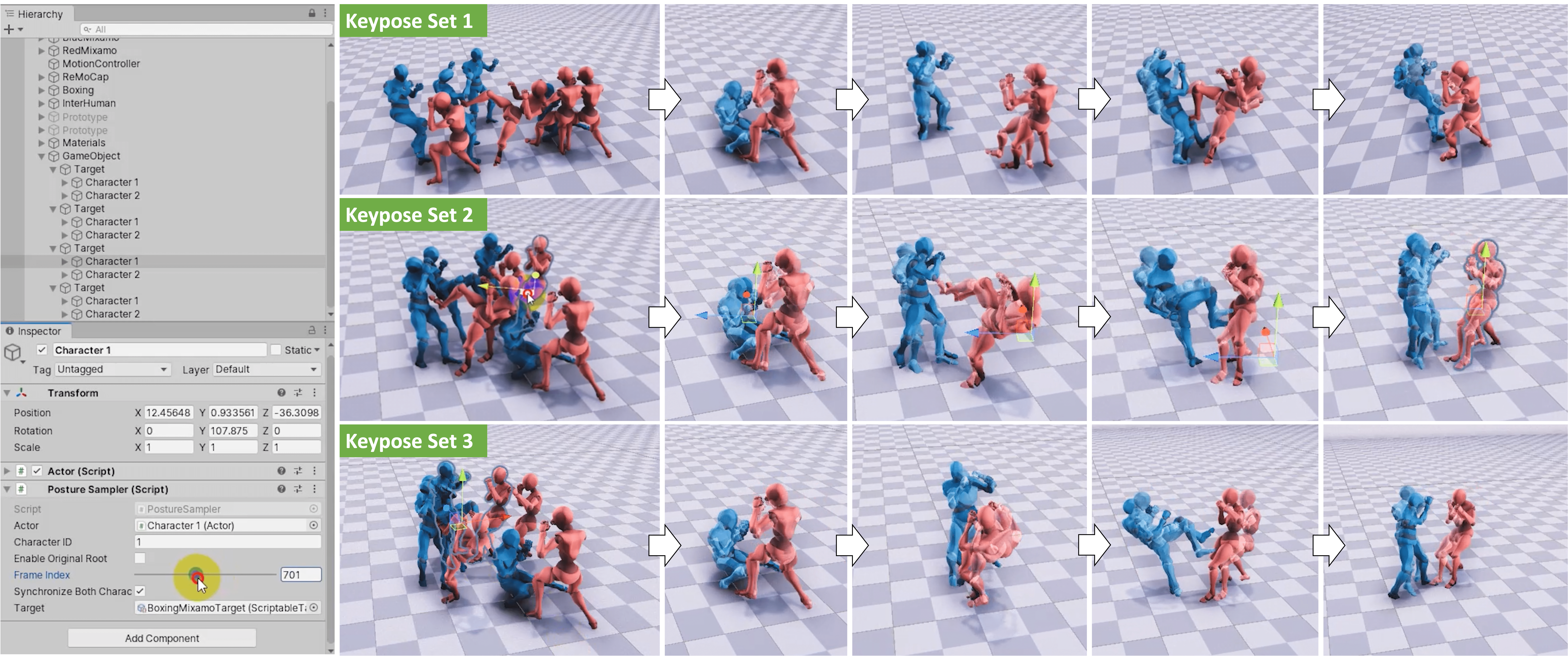}
  \caption{By manipulating the root translations and rotations of keyposes via the control panel (left), our system synthesizes interactive in-between motion sequences that dynamically respond to the specified keypose configurations. Detailed animation results are provided in the supplementary video.}
  \label{fig:ControlKeypose}
\end{figure*}

\end{document}